\newcounter{myctr}

\documentclass{myarticleclass}

\newcommand{\sa}{\sigma}			
\newcommand{\Sa}{\Sigma}			
\newcommand{\Pb}{\operatorname{P}}	
\newcommand{\Sh}{\operatorname{H}}	
\newcommand{\I}{\operatorname{I}}		
\newcommand{\Sm}{\tilde{S}}			
\newcommand{\sm}{\tilde{s}}			

\begin{document}

\markboth{Olof G\"{o}rnerup and Martin Nilsson Jacobi}{A Method for Inferring Hierarchical Dynamics in Stochastic Processes}

%
\catchline{}{}{}{}{}
%

\title{A METHOD FOR INFERRING HIERARCHICAL DYNAMICS \\ IN STOCHASTIC PROCESSES}

\author{\footnotesize OLOF G\"{O}RNERUP and MARTIN NILSSON JACOBI}
\address{Complex Systems Group, Department of Energy and Environment, \\
Chalmers University of Technology, 412 96 G\"{o}teborg, Sweden\\
olofgo@chalmers.se, mjacobi@chalmers.se}

\maketitle


\begin{abstract}
Complex systems may often be characterized by their hierarchical dynamics. In this paper do we present a method and an operational algorithm that automatically infer this property in a broad range of systems; discrete stochastic processes. The main idea is to systematically explore the set of projections from the state space of a process to smaller state spaces, and to determine which of the projections that impose Markovian dynamics on the coarser level. These projections, which we call \emph{Markov projections}, then constitute the hierarchical dynamics of the system. The algorithm operates on time series or other statistics, so a priori knowledge of the intrinsic workings of a system is not required in order to determine its hierarchical dynamics. We illustrate the method by applying it to two simple processes; a finite state automaton and an iterated map.
\end{abstract}

\keywords{Hierarchical dynamics; Model reduction; Coarse graining.}

\section{Introduction}

Modularity and hierarchical organization play an important role when determining the character of a dynamical system. It could even be argued that hierarchical self-organization is a necessary condition for a system to display a high degree of complexity, see e.g. Simon \cite{Simon62}. Hierarchical dynamics is also a prerequisite for efficient model reduction. Then the general strategy when reducing the level of details in a model is to find a partition of the degrees of freedom (i.e. a projection of the phase space) which by itself form a system with Markovian dynamics; an observation which is also discussed in detail by Shalizi and Moore \cite{Cosma}. Conversely one can use the idea behind the time-delay embedding method for attractor reconstruction \cite{Takens, Packard} to convince oneself that a dynamical system without the Markov property should be reconstructed in a higher dimensional phase space in order to make sense as a causal model.

In a physical system, modularity is usually associated with separations in time and length scales. In this paper do we focus on hierarchical decomposition of stochastic processes that, in general, are systems without direct physical interpretation. Despite the void of guidance from physical intuition, there are several methods that can be used for determining their hierarchical structure. For example, if one has full access to the inner workings of a process' dynamics,  i.e. the generative semi-group, Krohn-Rhodes theory can be used to decompose the semi-group as a hierarchical Wreath product of finite groups and finite aperiodic semi-groups \cite{ATOM}. Here do we present a method for hierarchical decomposition and reconstruction that instead operates on the sequence of states, i.e. a time series, that is generated by the process at hand. In this way, prior knowledge of the process' intrinsic dynamics is not necessary.

\subsection{Historical background}

The line of ideas on decomposition of dynamical systems and identification of hierarchical dynamics can be traced back to the analysis of continuous symmetries in classical mechanics as advanced by Lie\footnote{Strictly speaking, Lie's work is not limited to classical mechanics and can be applied to any differential equation.}, Lagrange, Poisson, Jacobi and Noether. These reduction schemes result in the elimination of inactive, i.e. constant, degrees of freedom. In non-equilibrium statistical mechanics, dimensional reduction generally means going from an effectively deterministic model to a Langevin type model that includes a noise term. The noise stems from fast, usually chaotic, motion that on the time scale of the relevant (slow) degrees of freedom can be approximated as white noise, resulting in a Markovian dynamics for the slow degrees of freedom. This idea was first formalized by Zwanzig \cite{zwanzig} and has later matured into the theory of adiabatic elimination, see e.g. \cite{Gardiner,Risken}. Yet another situation frequently encountered in models of natural systems is dissipative driven processes. A generic feature of such systems is that fast degrees of freedom, due to large negative Lyaponov exponents associated with the  dissipation, often relaxes to a quasi-fixed point, i.e. a point in the phase space that  appears effectively fixed on the time scale of the fast dynamics, but changes on the time scale set by the slow degrees of freedom. The overall dynamics is therefore in this case slaved to a slow positively invariant manifold and the resulting dimensionality is reduced. This picture has been advanced by Haken in his work on self-organization \cite{Haken}. Lately this idea has also been revitalized in the turbulence community, primarily by a proof of existence of inertial manifolds in a class of hyperbolic dynamical systems \cite{Foias}. In the same spirit, positive invariant manifolds are used in model reduction schemes in chemical kinetics \cite{Gorban}. Finally it is also worth mentioning that the connection between chaotic dynamical systems and non-equilibrium statistical mechanics has recently been further clarified by the work of Ruelle et. al. \cite{Ruelle,Dorfman}. 

\subsection{The method}

\begin{figure}
\begin{center}
\includegraphics[scale=0.22]{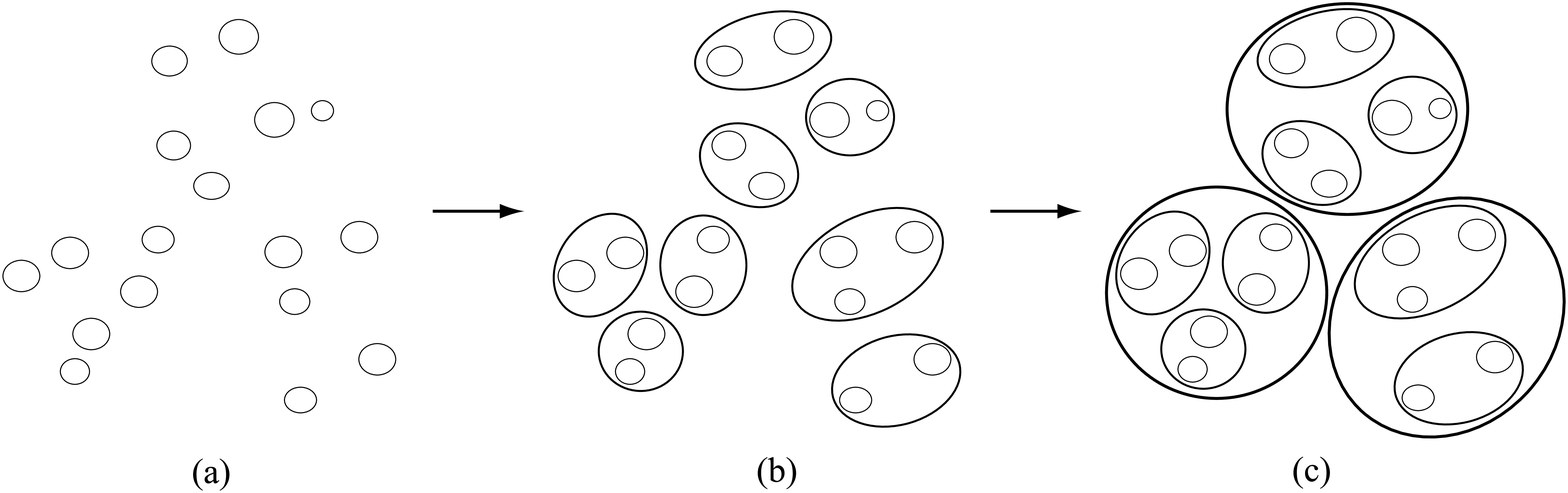}
\caption{Three modes of a dynamical system. A schematic illustration of how recursive modularity builds up hierarchies. There are three interacting modules in (c) that each contains interacting modules from (b) that each in turn contains interacting modules from (a). \emph{Given the dynamics of the modules in \emph{(a)}, how can we derive the existence and dynamics of the larger modules in \emph{(b)} and \emph{(c)}?}}
\label{modules}
\end{center}
\end{figure}

We start with a discussion on the general problem of how to define a hierarchical organization in a dynamical system. From a constructive point of view, a hierarchical system should be composed of interacting modules that contain, in some sense, smaller interacting modules. The process could be repeated to recursively generate new hierarchical levels as illustrated in Figure \ref{modules}. Conversely, a system is said to have hierarchical structure if it can be deconstructed through recursive decomposition of modular components.

\begin{figure}
\begin{center}
\includegraphics[scale=0.8]{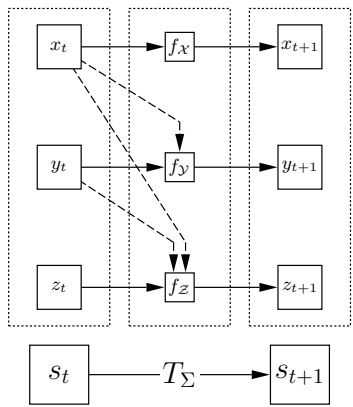}
\caption[]{A transition $s_t \mapsto s_{t+1}=T_{\Sa}(s_t)$ is decomposed into a transition in the Wreath product $(\mathcal{X}, T_{\mathcal{X}}) \wr (\mathcal{Y}, T_{\mathcal{Y}}) \wr (\mathcal{Z}, T_{\mathcal{Z}})$. The state $(z_t, y_t, x_t)$ is mapped to $(z_{t+1}, y_{t+1}, x_{t+1})=(z_t \cdot f_{\mathcal{Z}}(y_t, x_t), y_t \cdot f_{\mathcal{Y}}(x_t), x_t \cdot f_{\mathcal{X}})$ by the transformation $(f_{\mathcal{Z}}, f_{\mathcal{Y}}, f_{\mathcal{X}})$,  where $f_{\mathcal{Z}}: \mathcal{Y} \times \mathcal{X} \rightarrow \mathcal{Z}$, $f_{\mathcal{Y}}: \mathcal{X} \rightarrow \mathcal{Y}$ and $f_{\mathcal{X}} \in T_{\mathcal{X}}$, and where dot denotes group operation. See \cite{AHDOFSA} for further details. We may project away $z_t$ as $x_t$ and $y_t$ evolves independently of $z_t$, and we may project away $y_t$ \emph{together with} $z_t$ since they are slaved by $x_t$. We may not, on the other hand, project away $y_t$ by itself since it influences the evolution of $z_t$.}
\label{WreathTransition}
\end{center}
\end{figure}

\begin{figure}
\begin{center}
\includegraphics[scale=1.2]{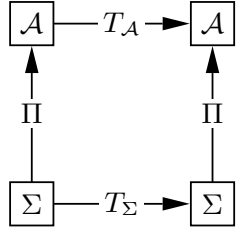}
\caption{An original state space $\Sigma$ is reduced to a coarser state space $\mathcal{A}$. The map $\Pi$ is a reduction. $T_{\Sigma}$ denotes the original update dynamics and $T_{\mathcal{A}}$ is the dynamics induced on $\mathcal{A}$ by $\Pi$. If $T_{\mathcal{A}}$ is a Markovian dynamics, the diagram commutes and we say that $\Pi$ is a Markov projection. }
\label{commuting_diagram}
\end{center}
\end{figure}

\begin{figure}
\begin{center}
\includegraphics[scale=0.9]{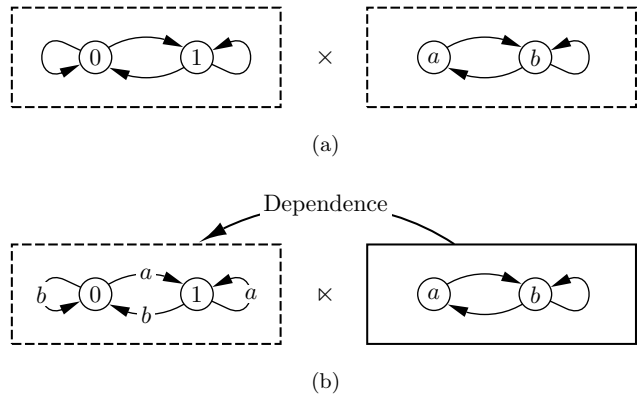}
\caption{(a) Direct product. Both automata can be projected away as they act independently of each other. (b) Semidirect product. Only the left automaton can be projected away since it is dependent on, i.e. is slaved by, the right automaton.}
\label{ProjAway}
\end{center}
\end{figure}

It is natural to invoke an operation that, in a categorical sense, carries the structure of a product: the Wreath product, see Figure \ref{WreathTransition}. Associated with the Wreath product, there exists a quotient operator, or a projection, which in essence defines divisibility of the system. The projection is in general a map from the full state space (phase space in the continuous case) to a smaller\footnote{Lower dimensionality in the continuous case and lower cardinality in the finite discrete case.} state space.\footnote{Normally a projection is also required to be idempotent, i.e. fulfill $P^2 = P$. In our case however, the projection maps between two different spaces and idempotency is not well defined.}  The dynamics of the original system  induces a dynamics on the projected state space. In principle any reducing map can be used as a hypothetical projection. However, crucial properties of the original dynamics are often lost in an arbitrary reduction. The most important such property is  the Markov property\footnote{Note that determinism is a special case of the Markov property.}. Only very special reductions, which we call {\em Markov projections}, do respect the fundamental character of the dynamics. When this happens, the diagram in Figure \ref{commuting_diagram} commutes. We may say that the dynamics is divisible by the quotient used by the projection, e.g. in Figure \ref{ProjAway}. For a continuous deterministic dynamical system we call a projection that respects the dynamics fiber preserving, although in this paper we focus on dynamics generated by discrete stochastic processes. The idea behind the method is to systematically test different reductions to see if they result in a Markovian dynamics on the reduced state space. If a reduction passes the test we conclude that it is a proper projection of the state space, i.e. a Markov projection.

The method introduced in this paper is inspired by computational mechanics \cite{ISC, CMPPSS} that derives optimal predictors of stochastic processes.\footnote{The idea of optimal predictors has been introduced on many different occasions, within different contexts. See \cite{CosmasThesis} for references and details.} The predictors, termed $\epsilon$-machines, are automata whose nodes are equivalence classes, termed causal states, of observed histories of states (i.e. semi-infinite sequences in the context of stochastic processes) such that two histories are equivalent if they condition the same probability distribution of future observed states. Causal states are connected with transitions that are labeled with the current state of the observed process, which completes the $\epsilon$-machine. An $\epsilon$-machine is the minimal and maximally efficient model of the observed process \cite{CMPPSS} and may in practice be acquired approximately e.g. from generated time series \cite{ShaliziEtAl04}. $\epsilon$-Machines are, in addition, Markov and one may use them to infer hierarchical dynamics in terms of causal states.

The dynamics represented by an $\epsilon$-machine operates on the raw micro state space of the observed process. It is therefore only a reduction if the original system has less active states than the state space admits. Our method does, in contrast, infer coarse grained dynamics on different hierarchical levels. Since inferred dynamics in our case by definition exhibits the Markov property, its finite state automaton representation is the minimal one and equivalent, subject to converting nodes to transitions\footnote{For the reader inclined to compare our approach and computational mechanics in more detail, it may be useful---in order to avoid initial confusion---to note that the states of a stochastic process label the \emph{transitions} of an $\epsilon$-machine, whereas they label the \emph{nodes} in our automaton representation.}, to the $\epsilon$-machine of the same coarse grained dynamics.

\section{Markov projections}
We will now describe our method in more detail. Before continuing, the reader may want to consult the appendix for a brief review of the concepts that are central to our approach and to see our use of notation.

\subsection{General idea}

\begin{figure}
\begin{center}
\includegraphics[scale=1.1]{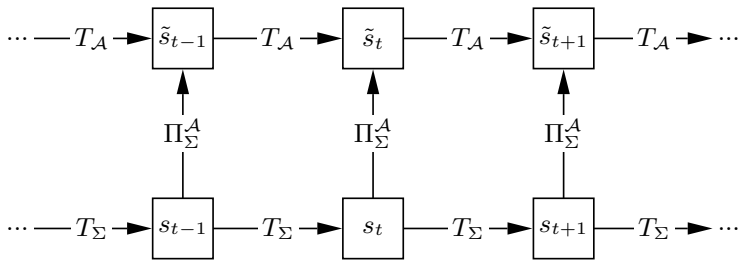}
\caption{The original symbol sequence $(..., s_{t-1}, s_t, s_{t+1}, ...)$ is projected onto a symbol sequence $(..., \sm_{t-1}, \sm_t, \sm_{t+1}, ...)$. Find a projection $\Pi_{\Sigma}^{\mathcal{A}}$ such that the imposed dynamics $T_{\mathcal{A}}$ has the Markov property. Then a state $\sm_{\tau}$ only depends on its previous state $\sm_{\tau-1}$.}
\label{SequenceDiagram}
\end{center}
\end{figure}

Say that you have a symbol sequence $s$ that has been generated by a stochastic process over some state space $\Sa$:
\begin{equation}
s=(..., s_{t-1}, s_t, s_{t+1}, ...), \ s_{\tau} \in \Sa.
\end{equation}
You wish to determine if  the process exhibits hierarchical dynamics according to the diagram in Figure \ref{commuting_diagram} and, if so, the nature of this hierarchy. To do this one may use a general procedure in the following. Systematically examine the set of possible partitions of $\Sa$. For each examined partition $\mathcal{A}$:
\begin{enumerate}
 \item Map $s$ onto a new symbol sequence $\sm$ and the corresponding process $T_{\mathcal{A}}$ with the projection $\Pi^{\mathcal{A}}_{\Sa} : \Sa \rightarrow {\mathcal A}$,
\begin{equation}
	\sm=(..., \sm_{t-1}, \sm_t, \sm_{t+1}, ...), \ \sm_{\tau}=\Pi^{\mathcal{A}}_{\Sa}(s_{\tau}),
\end{equation}

as in Figure \ref{SequenceDiagram}.
\item Test if the coarse grained sequence $\sm$ is constituted by Markovian dynamics.
\end{enumerate}
The Markov property in step (2) may be identified in the realm of information theory as we discuss next.

\subsection{Markov property measure}
Let $a_i$, $i=1, 2,..., |\mathcal{A}|$, be specific elements in a partition $\mathcal{A}$ (not to be confused with $s_{\tau}$, that are variables over elements in $\mathcal{A}$ at certain times $\tau$). For each state $a_i$, let $X_i$ be a stochastic variable of the past states preceding $a_i$, and let $Y_i$ be a stochastic variable of the subsequent state of $a_i$. If $X_i$ and $Y_i$ are independent for all $a_i \in \mathcal{A}$, $T_{\mathcal{A}}$ is a Markov process. There are a number of different ways to quantify the degree by which, or probability that, two distributions are independent. One common method is the $\chi ^2$ test \cite{NRIC}. Its main purpose, however, is to provide the significance of association between two variables, rather than the strength of association that we prefer. Although there indeed are measures of the strength of association based on $\chi ^2$ statistics, e.g. Cramer's $V$ and the contingency coefficient $C$ \cite{NRIC}, these are ill-suited for our purposes as the former exhibits discontinuities with varying contingency table size, and as the latter requires tables with an equal number of rows as columns. In addition do the measures lack direct interpretations. Instead do we use the mutual information $\I(X_i; Y_i)$ between $X_i$ and $Y_i$ as a measure of dependence with respect to state $a_i$. This measure grants clear-cut interpretations and, naturally, enables us to use tools from information theory. The mutual information is defined as 
\begin{equation}
\I(X_i; Y_i)=\Sh(X_i)+\Sh(Y_i)-\Sh(X_i, Y_i),
\label{MutInfoEq}
\end{equation}
where $\Sh(V)$ is the Shannon entropy
\begin{equation}
\Sh(V)=- \sum_{v \in \mathcal{V}} \Pb(V=v) \log_2 \Pb(V=v)
\end{equation}
of a stochastic variable $V$ drawn from $\mathcal{V}$, and where $\Sh(U, V)$ is the analogous entropy of the joint distribution of two stochastic variables $U$ and $V$. $\I(X_i; Y_i)$ is the information one gains when replacing the separate distributions $\Pb(X_i)$ and $\Pb(Y_i)$ with their joint distribution $\Pb(X_i, Y_i)$. $\I(X_i; Y_i) \geq 0$ with equality when $X_i$ and $Y_i$ are independent \cite{EIT}.
As a Markov property measure of a partition as a whole, we employ the expected mutual information
\begin{equation}
\langle \I \rangle = \sum_{a_i \in \mathcal{A}} \Pb(a_i) \I(X_i; Y_i),
\label{ExpIEq}
\end{equation}
where $\Pb(a_i) $ is probability of state $a_i$. Similarly, we use the shorthand $\Pb(..., \sm_{t-1}, \sm_{t}, \sm_{t+1})$ to denote the probability that a sequence of stochastic variables $(..., \Sm_{t-1}, \Sm_{t}, \Sm_{t+1})$  has the outcome $(..., \sm_{t-1}, \sm_{t}, \sm_{t+1})$. Using the definition for conditional probabilities and explicitly representing past and futures as substrings, we can rewrite Eq. \ref{ExpIEq} to
\begin{eqnarray}
\langle \I \rangle & = & - \sum_{\sm_t, \sm_{t+1}}\Pb(\sm_t, \sm_{t+1}) \log_2 \Pb(\sm_t, \sm_{t+1}) + \sum_{\sm_t}\Pb(\sm_t) \log_2 \Pb(\sm_t) \nonumber \\
& & -\sum_{..., \sm_{t-1}, \sm_t}\Pb(..., \sm_{t-1}, \sm_t) \log_2 \Pb(..., \sm_{t-1}, \sm_t) \nonumber \\ 
& & +\sum_{..., \sm_{t-1}, \sm_t, \sm_{t+1}}\Pb(..., \sm_{t-1}, \sm_t, \sm_{t+1}) \log_2 \Pb(..., \sm_{t-1}, \sm_t, \sm_{t+1})  \nonumber \\
& = & \Delta H_2 - \Delta H_{\infty},
\label{BlockEq}
\end{eqnarray}
where $\Delta H_n$ is the slope of the block entropy of $\Sm$ at length $n$ \cite{Shannon1948}. The expected mutual information $\langle \I \rangle$ between past symbols and the next symbol is in other words equivalent to the difference in expected uncertainty of a symbol conditioned on the preceding symbol and the expected uncertainty of a symbol conditioned on all preceding symbols. $\langle \I \rangle=0$ if one expects no reduction in the uncertainty of the current state from looking further back than one state.

In practice one acquires approximations of $X_i$ and $Y_i$ from a finite symbol sequence and finite history lengths. For each symbol $a_i \in \mathcal{A}$, we set up a contingency table whose rows are values of $X_i$ of occuring histories $(\sm_{t-n}, ..., \sm_{t-2}, \sm_{t-1})$ of length $n$ (drawn from $\mathcal{A}^n$) that precede $a_i$, and whose columns are values of $Y_i$ of possible subsequent symbols $\sm_t$ (drawn from $\mathcal{A}$) of $a_i$. Element $(j, k)$ in the table is then the count that history $j$ (according to some indexing) is followed by $a_i$ and then $a_k$. Eq. \ref{BlockEq} generalizes to finite history lengths:
\begin{equation}
\langle \I_n \rangle=\Delta H_2 - \Delta H_{n+2},
\end{equation}
where $\langle \I_n \rangle$ is the  expected mutual information when histories of length $n$ are considered.

\section{Examples}

Before moving on to algorithmic details, we exemplify our method by employing it to two simple stochastic processes; a finite state automaton and an iterated map.

\subsection{A simple automaton}

\begin{figure}
\begin{center}
\scalebox{1.0}{\includegraphics{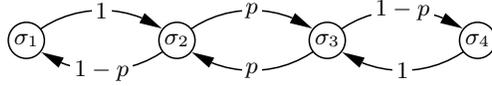}}
\end{center}
\caption{Example process $T_{\Sa}$ over the state space $\Sa=\{\sa_1, \sa_2,\sa_3,\sa_4\}$. Edges are labeled with transition probabilities.}
\label{ExProcOrig}
\end{figure}

\begin{figure}
\begin{center}
\scalebox{1.0}{\includegraphics{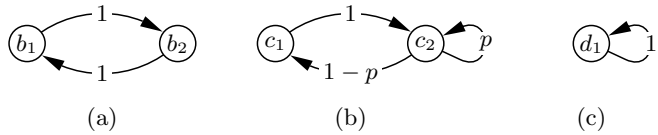}}
\end{center}
\vspace*{-8pt}
\caption{Dynamics (a) $T_{\mathcal{B}}$, (b) $T_{\mathcal{C}}$ and (c) $T_{\mathcal{D}}$ resulting from Markov projections of  the example process $T_{\Sa}$ in Figure \ref{ExProcOrig}.}
\label{ExProcFact}
\end{figure}

\begin{figure}
\begin{center}
\includegraphics[scale=0.9]{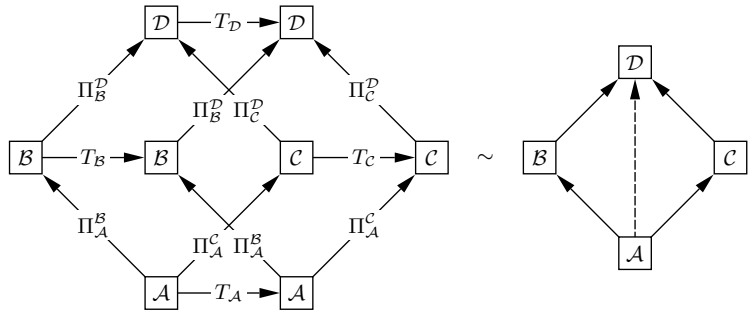}
\caption{Dynamics hierarchy of $T_{\Sa}$ for $p=1/2$ with simplified representation to the right. The dashed arrow implies that $\mathcal{A}$ may be directly projected onto $\mathcal{D}$. The arrow is redundant though, since projections may always be composed, e. g., $\Pi_{\mathcal{A}}^{\mathcal{D}} = \Pi_{\mathcal{B}}^{\mathcal{D}} \circ \Pi_{\mathcal{A}}^{\mathcal{B}}$ in this case.}
\label{AutomHierarchy}
\end{center}
\end{figure}

\begin{figure}
\begin{center}
\hspace{-40pt}
\scalebox{0.75}{\includegraphics{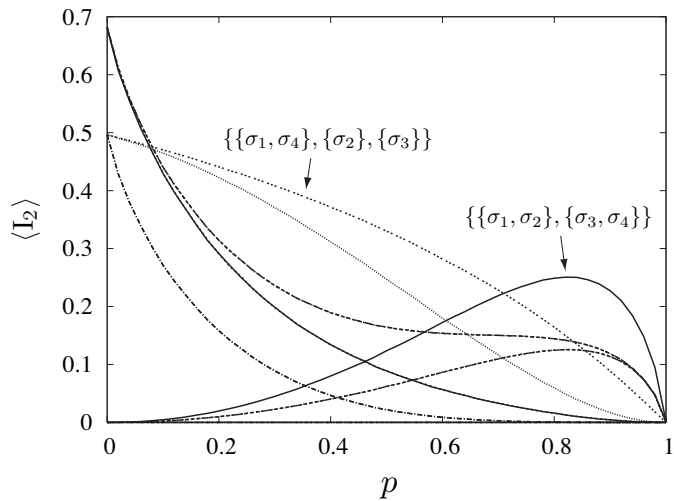}}
\end{center}
\vspace*{-8pt}
\caption{Expected mutual information in bits with respect to the 15 projections (of which two examples are labeled) of the process $T_{\Sigma}$ in Figure \ref{ExProcOrig} as functions of the transition probability $p$. Note that $\langle \I_2 \rangle$ of the Markov partitions is not visible since it is of the order of $10^{-6}$ bits. The statistics was collected from 1000 symbol sequences, each of length 1000.}
\label{IExpVsP}
\end{figure}

Consider the automaton $T_{\Sa}$ over the state space $\Sa=\{\sa_1, \sa_2,\sa_3,\sa_4\}$ in Figure \ref{ExProcOrig}. All possible 15 partitions are evaluated. For $p=1/2$, there are four Markov projections, including those corresponding to the trivial partitions $\mathcal{A}=\{\{\sa_1\},\{\sa_2\},\{\sa_3\},\{\sa_4\}\}$ and $\mathcal{D}=\{d_1\}$, where $d_1=\Sa$. The third process, $T_{\mathcal{B}}$, is a \emph{bit-flip} process over $\{b_1, b_2\}$, where $b_1=\{\sa_1, \sa_3\}$ and $b_2=\{\sa_2, \sa_4\}$, i.e.  repetition of $(b_1, b_2)$ blocks. The fourth projection gives $\mathcal{C}=\{c_1, c_2\}$, where $c_1=\{\sa_1, \sa_4\}$ and $c_2=\{\sa_2, \sa_3\}$. $T_{\mathcal{C}}$ is such that $c_1$ and $c_2$ are generated with probability $1-p$ and $p$ respectively, and a $c_1$ is always followed by a $c_2$. For $p=1/2$, this process is referred to as the \emph{golden mean process} \cite{RURO}. See Figure \ref{ExProcFact}. The Markov projections are related according to the hierarchy in Figure \ref{AutomHierarchy}, where the original process $T_{\Sa}$ is the direct product of $T_{\mathcal{B}}$ and $T_{\mathcal{C}}$; $T_{\Sa} = T_{\mathcal{B}}\times T_{\mathcal{C}}$. The number of Markov projections is dependent on the transition probability $p$. Figure \ref{IExpVsP} shows $\langle \I_2 \rangle$ for the separate possible partitions as functions of $p$. Specifically, at $p=0$ and $p=1$, more than four Markov projections exist (e.g. $\{\{\sa_1, \sa_4\},\{\sa_2\},\{\sa_3\}\}$ and $\{\{\sa_1, \sa_2\},\{\sa_3,\sa_4\}\}$) due to the elimination of transitions in  $T_{\Sa}$.

\subsection{An iterated map}

\begin{figure}

\begin{center}
\hspace{-40pt}
\scalebox{0.4}{\includegraphics{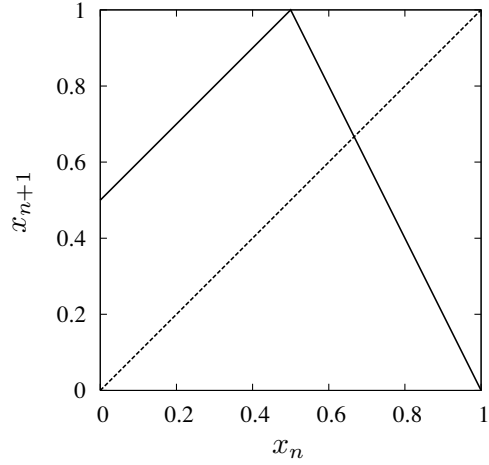}}
\end{center}
\caption{Roof map for $\alpha=1/2$.}
\label{roofMap}
\end{figure}

\begin{figure}
\begin{center}
\scalebox{1.0}{\includegraphics{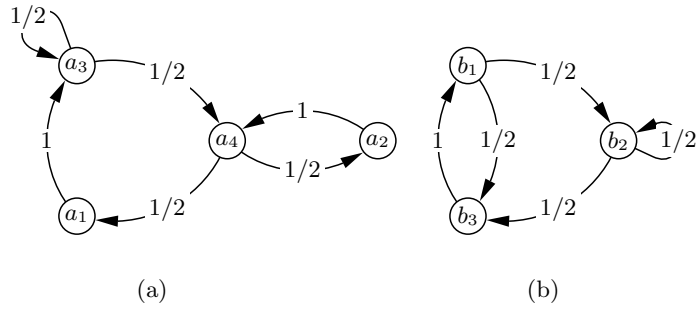}}
\end{center}
\vspace*{-8pt}
\caption{Markovian dynamics (a) $T_{\mathcal{A}}$ and (b) $T_{\mathcal{B}}$ of discretized Roof map for $\alpha=1/2$.}
\label{roofProjDyn}
\end{figure}

\begin{figure}[htbp]
\begin{center}
\includegraphics[scale=0.9]{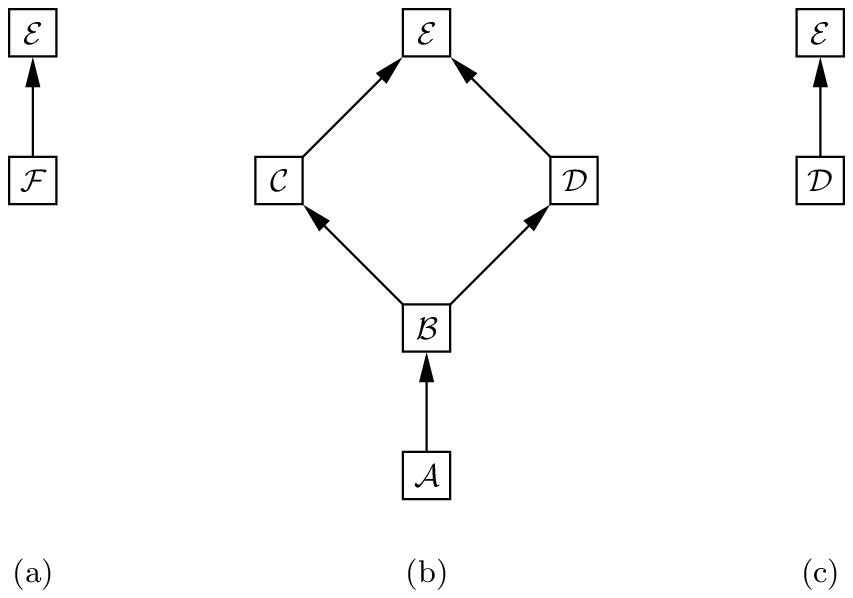}
\caption{Dynamics hierarchies of discretized Roof map for (a) $\alpha=1/4$, (b) $\alpha=1/2$ and (c) $\alpha=3/4$. Dashed arrows for composed projections are omitted.}
\label{RoofHierarchy}
\end{center}
\end{figure}

The second example concerns a process that belongs to a well-studied class of dynamical systems; iterated maps. They are discrete in time, but operate on continuous phase spaces. Our process, the \emph{Roof map}, is defined over $[0, 1]$ according to
\begin{equation}
x_{n+1}=
\begin{cases}
\frac{1- \alpha}{\alpha}x_n +\alpha, & \text{if $x_n < \alpha$} \\
\frac{x_n-1}{\alpha-1} & \text{otherwise},
\end{cases}
\label{RoofMapEq}
\end{equation}
where $\alpha$ is a parameter.  We iterate the map and discretize the trajectory with four equally large bins. That is, $x_i \mapsto \sa_1$ if $x_i \in [0, 1/4[$, $x_i \mapsto \sa_2$ if $x_i \in [1/4, 1/2[$, $x_i \mapsto \sa_3$ if $x_i \in [1/2, 3/4[$ and $x_i \mapsto \sa_4$ otherwise. For $\alpha=1/2$, Figure \ref{roofMap}, there are five Markov projections that  correspond to the partitions 
\begin{eqnarray}
\mathcal{A}&=&\{\{\sa_1\}, \{\sa_2\}, \{\sa_3\}, \{\sa_4\}\}, \nonumber \\
\mathcal{B}&=&\{\{\sa_1 , \sa_2\}, \{\sa_3\}, \{\sa_4\}\}, \nonumber \\
\mathcal{C}&=&\{\{\sa_1, \sa_2\}, \{\sa_3, \sa_4\}\}, \nonumber \\
\mathcal{D}&=&\{\{\sa_1, \sa_2, \sa_3\}, \{\sa_4\}\} \text{ and}  \nonumber \\
\mathcal{E}&=&\{\{\sa_1, \sa_2, \sa_3, \sa_4\}\}. \nonumber
\end{eqnarray}
$T_{\mathcal{A}}$ and $T_{\mathcal{B}}$ are shown in Figure \ref{roofProjDyn}, whereas $T_{\mathcal{C}}$ and $T_{\mathcal{D}}$ both are isomorphic to the automaton in Figure \ref{ExProcFact}(b) with $p=1/2$. The Markov projections are related according to the hierarchy in Figure \ref{RoofHierarchy}(b). Note that $\mathcal{E}$ is the one-element partition that trivially implies Markovian dynamics. For $\alpha=1/4$ and $\alpha=3/4$, Figure \ref{RoofHierarchy}(a) and \ref{RoofHierarchy}(c) respectively, the Markovian dynamics of finest resolution is not provided by $\mathcal{A}$. Instead, the partitions 
\begin{eqnarray}
\mathcal{F}&=&\{\{\sa_1\}, \{\sa_2, \sa_3, \sa_4\}\}, \nonumber
\end{eqnarray}
for $\alpha=1/4$, and $\mathcal{D}$, for $\alpha=3/4$, are required, where $T_{\mathcal{F}}$ and $T_{\mathcal{D}}$ are isomorphic.

\section{Algorithm}
\label{AlgorithmSec}

Although an exhaustive search through the set of all possible partitions works fine for small $|\Sa|$, it is an impractical strategy for larger state spaces due to a combinatorial explosion. A state space with $n$ states allows for $B_n = \sum _{k=1} ^n S(n,k) $ partitions, where $B_n$ is called the Bell number and $S(n,k)$ is the Stirling number. The latter is the number of ways to partition a set with cardinality  $n$ into $k$ nonempty subsets. Both $B_n$ and $S(n,k)$ are large numbers (assuming $1<k<n-1$),
\begin{eqnarray*}
            S (n,k )  & = & \frac{1}{k!} \sum _{j=1} ^k (-1) ^{k-j} \binom{k}{j}  j ^n,
\end{eqnarray*}
which calls for an approach other than sheer brute force.

\subsection{Recursive partitions}
There are some basic relations between partitions that allow us to avoid an exhaustive search of all possible partitions. When designing our algorithm, we exploit  that the mutual information of a partition $\mathcal{A}$, with respect to a state $a_i$, is larger or equal to the mutual information of a partition $\mathcal{B} \ni a_i$ of $\mathcal{A}$, with respect to $a_i$. To see this inequality, consider the following. Let $\Pi$ be a projection from $\mathcal{A}$ to $\mathcal{B}$; $X_i$ and $\tilde{X}_i$ stochastic variables of the past states preceding $a_i$ with respect to $\mathcal{A}$ and $\mathcal{B}$; and $Y_i$ and $\tilde{Y}_i$ stochastic variables of the subsequent states of $a_i$ with respect to $\mathcal{A}$ and $\mathcal{B}$. Since $Y_i$ is a function whose range is $\mathcal{A}$, it may be composed with $\Pi$. Then 
\begin{equation}
\Pi(Y_i)=\tilde{Y}_i.
\end{equation}
Let $\Phi$ be a function that maps semi-infinite sequences of states from $\mathcal{A}$ onto semi-infinite sequences of states from $\mathcal{B}$,
\begin{equation}
\Phi(..., s_{t-3}, s_{t-2}, s_{t-1})=(..., \Pi(s_{t-3}), \Pi(s_{t-2}), \Pi(s_{t-1})).
\end{equation}
Then
\begin{equation}
\Phi(X_i)=\tilde{X}_i.
\end{equation}
Since a function $g$ of a stochastic variable $V$ cannot increase the information about another stochastic variable $U$,  $\I(U; V) \geq \I(U; g(V))$ \cite{EIT} (p. 35), we see that
\begin{equation}
\I(X_i; Y_i) \geq \I(X_i; \Pi(Y_i)) \geq \I(\Phi(X_i); \Pi(Y_i))= \I(\tilde{X}_i; \tilde{Y}_i),
\label{IInequality}
\end{equation}
as $\I$ is symmetric.
The inequality (\ref{IInequality}) is helpful since if we know that an element $b_i$ in a partition $\mathcal{B}$ results in high mutual information, we may discard all partitions $\mathcal{A} \ni b_i$ that projects onto $\mathcal{B}$. This leads us to an algorithm that evaluates partitions in ascending cardinal order, i.e. from coarser to finer partitions.

\subsection{Procedure}
The components of the algorithm are the sets $\mathcal{S}_p$ (previous partition elements) and $\mathcal{S}_c$ (current partition elements), and the integer $l$ (level). Left arrow ($\leftarrow$) denotes assignment. 
\begin{enumerate}
\item Initiation: $\mathcal{S}_p \leftarrow 2^{\Sa}$ (the power set of $\Sa$), $\mathcal{S}_c \leftarrow \emptyset$ and $l \leftarrow 2$.
\item For every partition of size $l$ composed from elements in $\mathcal{S}_p$
	\begin{enumerate} 
		\item Evaluate $\langle \I_n \rangle$ and store Markov projections. 
		\item Add elements of size $\leq  |\Sa|-l$ with low mutual information $\I_n$ to $\mathcal{S}_c$. 
	\end{enumerate} 
	If no partitions can be composed or if $l = |\Sa|$, stop.
\item $\mathcal{S}_p \leftarrow \mathcal{S}_c$, $\mathcal{S}_c \leftarrow \emptyset$ and $l \leftarrow l+1$. Go to step (2)
\end{enumerate}
Partitions of size $l$ that are evaluated are thus those that can be composed from elements that have implied low mutual information at level $l-1$. In such a way, partition elements that result in high mutual information are successively discarded as these may not improve due to Eq. \ref{IInequality}.

\subsection{Possible further pruning}

The algorithm may indeed be subject to improvements. If we assume that the dynamics is ergodic then it follows that the cardinality of the process as a whole, i.e. $| \Sigma |$, has the cardinality of its components as divisors. In other words, if we have reason to assume that the time series that we observe includes all possible states in the state space (in essence a weak ergodicity assumption), then we only need to try partitions with cardinality that divides $| \Sigma |$. The reason for this is straight forward. Assume that we combine two processes ${\mathcal T}_A$ and ${\mathcal T}_B$ with transition matrices $Q_A$ ($n \times n$) and $Q_B$ ($m \times m$) respectively. If the two processes are combined in a trivial way, i.e. there is no interaction between the sub-processes ${\mathcal T}_A$ and ${\mathcal T}_B$, then the total process has the transition matrix $Q = Q_A \otimes Q_B$, a $n \cdot m \times n \cdot m$ matrix. Our algorithm works from the other end. We are given a sequence of states from which we can estimate the total transition matrix. Then it is clear that the number of states in an independent sub-process must be a devisor of the number of states in the process as a whole.

More generally, two processes can be combined in a more non-trivial fashion where for example ${\mathcal T}_A$ is slaved by ${\mathcal T}_B$, i.e. the dynamics of ${\mathcal T}_A$ is affected by the state of ${\mathcal T}_B$ but not vice versa (e.g. as in Figure \ref{ProjAway}(b)). The resulting process is then described by a semi-direct product or a Wreath product. We will not go into the details of this algebraic structure here but refer the reader to any introductory text on group theory, for example \cite{MA}.  The overall conclusion above is however still valid; the number of stats in the process ${\mathcal T}_B$ must be a divisor of the number of states in the combined process. We can still  decrease the number of tested partitions, again, under the assumption that we have global coverage of the state space. 

\section{Discussion}

We have presented a method for inferring hierarchical dynamics from observed time series. Alternatively we may say that the presented scheme detects components of a process that in themselves have Markovian dynamics. The possible usefulness of this, as well as other related methods for decomposing and reducing dynamical systems, is great. Essentially all related methods are, however, either tailored for a limited class of systems, or suffers from high computational complexity. Examples of methods that exhibit the latter are Krohn-Rodes theory, calculations of invariant manifolds, and Markov partitions \cite{Remo99}. Our method is no exception though. The added structure introduced in Section \ref{AlgorithmSec} does indeed reduce the number of potential partitions of the state space, but it is not enough to remedy a computational cost that in the worst case scales exponentially with the cardinality of the state space. We believe that this problem is generic to any reduction scheme. In practice one must hope that the process under analysis carries some additional structure that allows us to make further assumptions about which type of projections that make sense to test. For example, if the system is a cellular automaton, it is clear that the reduction should be faithful to the locality and translational invariance of the update rule, i.e. only projections acting locally and independent of the position on the underlying lattice should be evaluated. The reduction of possible partitions from such considerations often recasts the problem into the computationally feasible domain. Note that the tailoring needed is limited to the generation of test partitions, the rest of the algorithm remains unchanged. This feature is appealing from the implementation standpoint.
 
We end this presentation by briefly mentioning the types of systems for which we hope that the method is useful. These may be spin systems (renormalization), lattice gases (automated detection of hydrodynamics variables), pattern forming cellular automata (it is speculated, and partially known, that these systems have hierarchies of descriptions), and interaction networks (identification of functional groups).
 
\section*{Acknowledgments}

The authors would like to thank Anders Eriksson for helpful discussions and comments. This work was funded by PACE (Programmable Artificial Cell Evolution), a European Integrated Project in the EU FP6-IST-FET Complex Systems Initiative, and by EMBIO (Emergent Organisation in Complex Biomolecular Systems), a European Project in the EU FP6-NEST Initiative.

\bibliography{mProj}

\appendix
\section{Preliminaries}

\subsection{Stochastic processes}
Here we consider dynamical systems in the form of discrete stochastic processes. They generate bi-infinite sequences $S$ of stochastic variables $S_{\tau}$,
\begin{equation}
S=(..., S_{t-1}, S_t, S_{t+1}, ...),
\end{equation}
where each $S_{\tau}$ is drawn from a state space $\Sa=\{\sa_1, \sa_2, ..., \sa_n\}$. The subscript $t$ denotes the present state, whereas $t+i$ with negative and positive indices $i$ denote past and future states, respectively. We always assume that a process that generates $S$ is \emph{stationary}, i.e. that it is invariant under time translation:
\begin{eqnarray}
& & P(S_{t+n}=s_0, S_{t+n+1}=s_1, ..., S_{t+n+l}=s_l)= \\
& & P(S_{t+m}=s_0, S_{t+m+1}=s_1, ..., S_{t+m+l}=s_l),
\end{eqnarray}
for all $n, m \in \mathbb{Z}$, $l \in \mathbb{N}$  and $s_k \in \Sa$.

\subsection{Markov processes}
A process is said to have the \emph{Markov property} if future states are conditionally independent of past states, given the current state:
\begin{eqnarray}
& & P(S_{t+n}=s_0 | ..., S_{t+n-3}=s_3, S_{t+n-2}=s_2, S_{t+n-1}=s_1) = \\
& & P(S_{t+n}=s_0 | S_{t+n-1}=s_1),
\end{eqnarray}
for all $n \in \mathbb{Z}$ and $s_k \in \Sa$. Processes with this property are termed \emph{Markov processes}.

\subsection{Partitions and projections}
\label{App3}
By a \emph{partition} $\mathcal{A}$ of the state space $\Sa$, we refer to a set of disjoint subsets whose union is $\Sa$:
\begin{equation}
\mathcal{A}=\{a_1, a_2, ..., a_n\},
\end{equation}
where $\bigcup_{i=1}^{n} a_i = \Sa$ and $a_i \cap a_j=\emptyset$ for all $i \neq j$. A \emph{projection}
\begin{equation}
\Pi^{\mathcal{A}}_{\Sa}: \Sa \rightarrow \mathcal{A}
\label{ProjEq}
\end{equation}
is a function that maps elements of $\Sa$ onto their respective elements in $\mathcal{A}$. 

Further, one may recursively partition a partition. A partition $\mathcal{B}$ of another partition $\mathcal{A}$ is a set of unions of disjoint subsets of $\mathcal{A}$, where the union of the elements of $\mathcal{A}$ and the union of the elements of $\mathcal{B}$ are equal. That is,
\begin{equation}
\mathcal{B}=\{b_1, b_2, ..., b_m\},
\end{equation}
where $b_i=\bigcup a_i,$  $a_i \in \mathcal{A}$; $b_i \cap b_j = \emptyset$ for all $i \neq j$, and  $\bigcup_{i=1}^{n} a_i = \bigcup_{j=1}^{m} b_j = \Sa$ .
The projection $\Pi^{\mathcal{B}}_{\mathcal{A}}$ from $\mathcal{A}$ to $\mathcal{B}$ is analog to (\ref{ProjEq}).

\subsection{Example}

We conclude the preliminaries with a simple example. Consider the process $T_{\Sa}$ given by the finite state automaton in Figure. \ref{ExProcOrig}. $T_{\Sa}$ acts on the alphabet $\Sa=\{\sa_1, \sa_ 2, \sa_3, \sa_4\}$, is stationary, and fulfills the Markov property since the transition probabilities from each state are independent of previously visited states. 
There are $15$ different possible partitions of $\Sa$; for example $\mathcal{A}=\{\{\sa_1\},\{\sa_2, \sa_3\}, \{\sa_4\}\}$, $\mathcal{B}=\{\{\sa_1\},\{\sa_2, \sa_3, \sa_4\}\}$, $\mathcal{C}=\{\{\sa_1\},\{\sa_2\},\{\sa_3\},\{\sa_4\}\}$ and $\mathcal{D}=\{\Sa\}$, where $\mathcal{B}$, e. g., is a partition of $\mathcal{A}$. The projection $\Pi^{\mathcal{A}}_{\Sa}$, for instance, gives $\Pi^{\mathcal{A}}_{\Sa}(\sa_1)=\{\sa_1\}$, $\Pi^{\mathcal{A}}_{\Sa}(\sa_2)=\Pi^{\mathcal{A}}_{\Sa}(\sa_3)=\{\sa_2, \sa_3\}$ and $\Pi^{\mathcal{A}}_{\Sa}(\sa_4)=\{\sa_4\}$.

\end{document}